\documentclass[twocolumn,twoside,slac_two]{revtex4}
\pdfoutput=1
\usepackage{graphicx}
\usepackage{fancyhdr}
\begin{document}

\title{Longitudinal EAS-Development Studies in the Air-Shower Experiment KASCADE-Grande}
%
%
\author{P. Doll$^{b \footnote{corresponding author, e-mail:
paul.doll@kit.edu}}$, W.D. Apel$^{b}$, J.C.
Arteaga-Vel\'azquez$^{a,\footnote{now at: Instituto de F\'\i sica y Matem\'aticas, Universidad Michoacana, Morelia, Mexico \\
$^{2}$ now at: Max-Planck-Institut f\"ur Physik, M\"unchen, Germany\\
$^{3}$ now at: Universidade S$\tilde{a}$o Paulo, Instituto de F\'\i sica de S$\tilde{a}$o Carlos, Brasil\\
$^{4}$ now at: Dept. of Astrophysics, Radboud University Nijmegen, The Netherlands\\
$^{5}$ now at: Institute for Space Sciences, Bucharest-Magurele, Romania\\
$^{6}$ deceased\\
$^{7}$ now at: University of Trondheim, Norway\\}} $,
K.Bekk$^{b}$,
M. Bertaina$^{c}$, J. Bl\"umer$^{a,b}$, H. Bozdog$^{b}$,\\
I.M. Brancus$^{d}$, P. Buchholz$^{e}$, E. Cantoni$^{c}$, A.
Chiavassa$^{c}$, F. Cossavella$^{a,2
}$, K. Daumiller$^{b}$,\\
V. de Souza$^{a,3}$, F. di Pierro$^{c}$, R. Engel$^{b}$, J.
Engler$^{b}$,
M. Finger$^{a}$, D. Fuhrmann$^{f}$, P.L. Ghia$^{g}$,\\
H.J. Gils$^{b}$, R. Glasstetter$^{f}$, C. Grupen$^{e}$, A.
Haungs$^{b}$, D. Heck$^{b}$, J.R. H\"orandel$^{a,4}$,
T. Huege$^{b}$,\\
P.G. Isar$^{b,5}$, K.-H. Kampert$^{f}$, D. Kang$^{a}$ D.
Kickelbick$^{e}$, H.O. Klages$^{b}$, K. Link$^{a}$, P.
{\L}uczak$^{h}$,
M. Ludwig$^{a}$,\\
H.J. Mathes$^{b}$, H.J. Mayer$^{b}$, M. Melissas$^{a}$, J.
Milke$^{b}$,
B. Mitrica$^{d}$, C. Morello$^{g}$, G. Navarra$^{c,6}$,\\
S. Nehls$^{b}$, J. Oehlschl\"ager$^{b}$, S. Ostapchenko$^{b,7}$,
S. Over$^{e}$, N. Palmieria$^{a}$,
M. Petcu$^{d}$, T. Pierog$^{b}$,\\
H. Rebel$^{b}$, M. Roth$^{b}$, H. Schieler$^{b}$, F.G.
Schr\"oder$^{b}$,
O. Sima$^{i}$, G. Toma$^{d}$, G.C. Trinchero$^{g}$,\\
H. Ulrich$^{b}$, A. Weindl$^{b}$, J. Wochele$^{b}$, M.
Wommer$^{b}$, J. Zabierowski$^{h}$}

\affiliation{ $^{a}$Institut f\"ur Experimentelle Kernphysik,
KIT - Campus S\"ud, 76021 Karlsruhe, Germany\\
$^{b}$Institut f\"ur Kernphysik, KIT -
Campus Nord, 76021 Karlsruhe, Germany\\
$^{c}$Diparimento di Fisica Generale dell'Universit\`a, 10125
Torino,
Italy\\
$^{d}$National Institute of Physics and Nuclear Engineering, 7690
Bucharest, Romania\\
$^{e}$Fachbereich Physik, Universit\"at Siegen, 57068 Siegen, Germany\\
$^{f}$Fachbereich Physik, Universit\"at Wuppertal, 42097
Wuppertal,
Germany\\
$^{g}$Istituto di Fisica dello Spazio Interplanetario, INAF, 10133
Torino,
Italy\\
$^{h}$Soltan Institute for Nuclear Studies, 90950 Lodz, Poland\\
$^{i}$Department of Physics, University Bucharest, 76900
Bucharest, Romania\\}

\begin{abstract}
A large area ($128m^{2}$) Muon Tracking Detector (MTD), located
within the KASCADE experiment, has been built with the aim to
identify muons ($E_{\mu}>$0.8GeV) and their directions in
extensive air showers by track measurements under more than 18
r.l. shielding. The orientation of the muon track with respect to
the shower axis is expressed in terms of the radial- and
tangential angles. By means of triangulation the muon production
height $H_{\mu}$ is determined. By means of $H_{\mu}$, a
transition from light to heavy cosmic ray primary particle with
increasing shower energy $E_{o}$ from 1-10 PeV is observed. Muon
pseudorapidity distributions for the first interactions above 15km
are studied and compared to Monte Carlo simulations. \vspace{1pc}

\end{abstract}

\maketitle
\thispagestyle{fancy}

\section{Introduction}

Muons have never been used up to now to reconstruct the hadron
longitudinal development of EAS with sufficient accuracy, due to
the difficulty of building large area ground-based muon telescopes
~\cite{doll01}. Muons are produced mainly by the decay of charged
pions and kaons in a wide energy range. They are not always
produced directly on the shower axis. Multiple Coulomb scattering
in the atmosphere and in the detector shielding may change the
muon direction. It is evident that the reconstruction of the
longitudinal development of the muon component by means of
triangulation ~\cite{{amb3},{obenland}} provides a powerful tool
for primary mass measurement ~\cite{doll} , giving an information
similar to that obtained with the Fly's Eye experiment, but in the
energy range not accessible by the detection of fluorescence
light. Muon tracking allows also the study of hadron interactions
by means of the muon pseudorapidity ~\cite{zabier}. Already in the
past, analytical tools have been developed which describe the
transformation between shower observables recorded on the ground
and observables which represent directly the longitudinal shower
development~\cite{Pent}. Fig. 1 in ref ~\cite{zabier1} shows the
experimental environment. Measured core position distributions for
showers inside KASCADE range from 40 - 120 m and inside Grande
from 250 - 360 m.The shower core position ranges cover full
trigger efficiency as confirmed by investigations of muon lateral
density distributions ~\cite{luczak}. With CR studies very high
energies are accessible in the 'knee' energy region $10^{15}$ -
$10^{16.5}$ eV, which correspond to CM energies in the
nucleon-nucleon system from 1.4 - 8 TeV currently covered by the
Tevatron and the LHC. Provided, that it is feasible to focus on
the first encounters with the atmospheric nuclei, muon
multiplicity studies may provide insight into the high energy
interaction at around 8 TeV. In this energy range saturation
physics is expected to enter (rise of $<p_{T}>$) which is
transformed to geometric scaling ~\cite{wolschin} by the
color-glass-condensate theory.

\section{Muon Production Height}

The angular correlation of the muon tracks with respect to the
shower axis is expressed by the $\rho$ and the $\tau$ angles
~\cite{doll01}. The $\rho$ angle contains some scattering which is
represented by the $\tau$ angle value exhibiting a $\sigma_{\tau}
\sim 0.2^{o}$.

 Fig. 1 shows $\rho$ angle distributions for specific muon
number $lg(N_{\mu})$ bins corresponding to different shower energy
bins ~\cite{doll}. The $\rho$ angle distributions are plotted for
'light' and 'heavy' primary CR mass enriched showers, employing
the $lg(N_{\mu})/lg(N_{e})$ ratio (corrected for attenuation)
~\cite{weber}~\cite{doll} to be larger ('heavy') or smaller
('light') than 0.83. The distributions show a dependence on the
primary mass range, however, masked by the energy dependent
penetration.

\begin{figure}
\begin{center}
\includegraphics [width=0.55\textwidth]{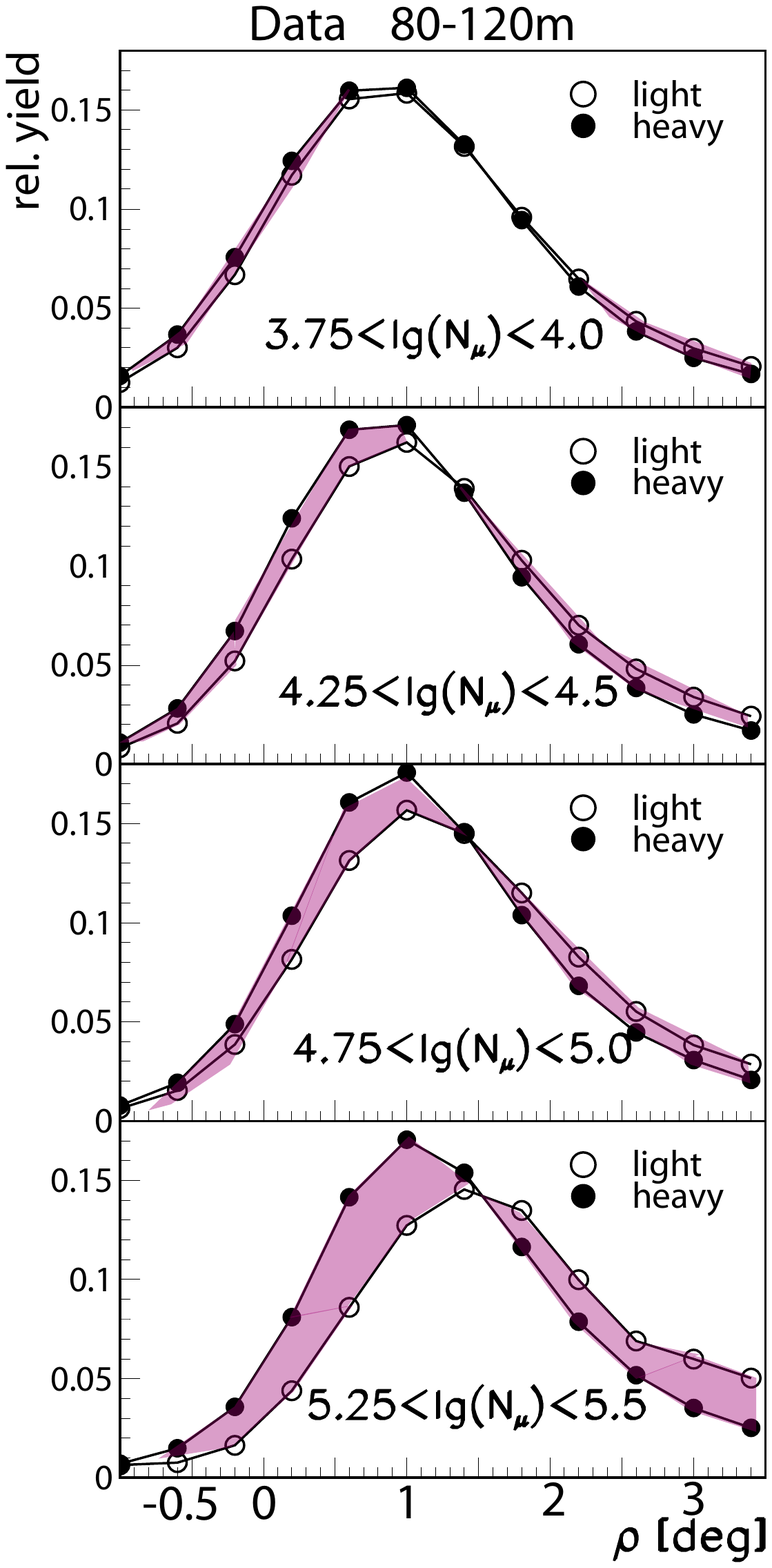}
\end{center}
\vspace{-2.0cm} \caption{  $\rho$ angle distributions (normalized
to integral yield equal to one) for different muon size bins and
different $lg(N_{\mu})/lg(N_{e})$ ratio above 0.83 ('heavy') and
below 0.83 ('light'), for 80 - 120 m core distance range and
shower direction $\Theta=0^{o}-18^{o}$. Lines connect points. }
\label{fig1}
\end{figure}


Based on $\rho$ and $\tau$ angles and the distance of the muon hit
to the shower core $R_{\mu}$, the muon production height $h_{\mu}$
along the shower axis is calculated:
\begin{equation}
h_{\mu}=R_{\mu}/\tan(\rho-\mid\tau\mid)
\end{equation}

The MTD-KASCADE system with its dense array grid and 80 - 120 m
core distance range for the muon track allows to study the muon
momenta in the $100-200~GeV$ range. $h_{\mu}$ will be considered
for $\rho>\tau$ which we can extend up to 20 km for a shower core
muon hit distance window 80 - 120 m (note $100~m/20000~m \cong
0.28^{o}$) employing, according to simulations, $\sim 200~GeV$
muons from above $15~km$.
 Fig. 2 shows muon production height distributions for
different muon size bins and different $lg(N_{\mu})/lg(N_{e})$
ratio above 0.83 ('heavy') and below 0.83 ('light'). CORSIKA
~\cite{heck} simulations based on QGSjetII+FLUKA2002.4 ~(slope
-2.7 and -3.1 below and above the knee, respectively) for Hydrogen
and Iron are shown in the Fig. 2 as well. In the low $h_{\mu}$
range the low energy interaction model (FLUKA2002.4) seems capable
to describe the $h_{\mu}$ distribution. The experimental
distributions are getting more narrow with increasing energy but
differently for 'light' and 'heavy' CR primaries. The $S_{NN}$
numbers quote the CM energies assuming A=1 CR primary. In the
course of the analysis in ~\cite{doll}, $h_{\mu}$ is transformed
to $H_{\mu}~[g/cm^{2}]$ and after subtracting from each $H_{\mu}$
an 'energy' dependent penetration depth the remaining depth
$H_{\mu}^{A}$ exhibits ~\cite{doll} the mass A sensitivity. Mass
resolution is limited and $<H_{\mu}^{A}> \sim <H_{\mu}^{H}>
-~30g/cm^{2} lnA$ ~\cite{doll}.

\begin{figure}
\begin{center}
\includegraphics [width=0.48\textwidth]{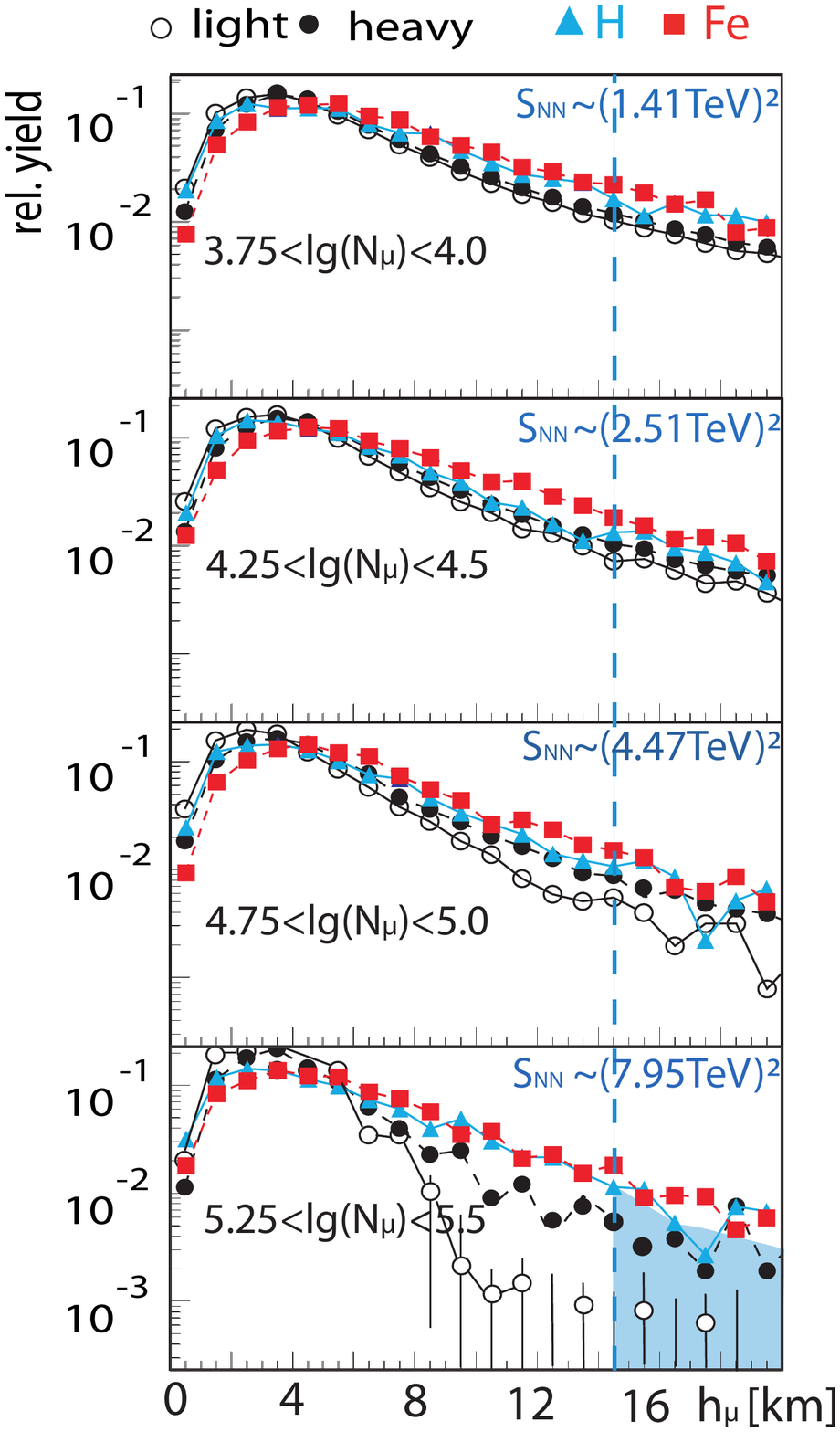}
\end{center}
\vspace{-1.cm} \caption{ Muon production height distributions
(normalized to integral yield equal to one) for different muon
size bins and different $lg(N_{\mu})/lg(N_{e})$ ratio above 0.83
('heavy') and below 0.83 ('light') and 80 - 120 m core distance
range and for shower direction $\Theta=0^{o}-18^{o}$. Also CORSIKA
simulations based on~QGSjetII + FLUKA2002.4 are given for Hydrogen
(H) and Iron (Fe) CR primaries. Lines connect points.}
\label{fig2}
\end{figure}

\begin{figure}
\begin{center}
\includegraphics [width=0.55\textwidth]{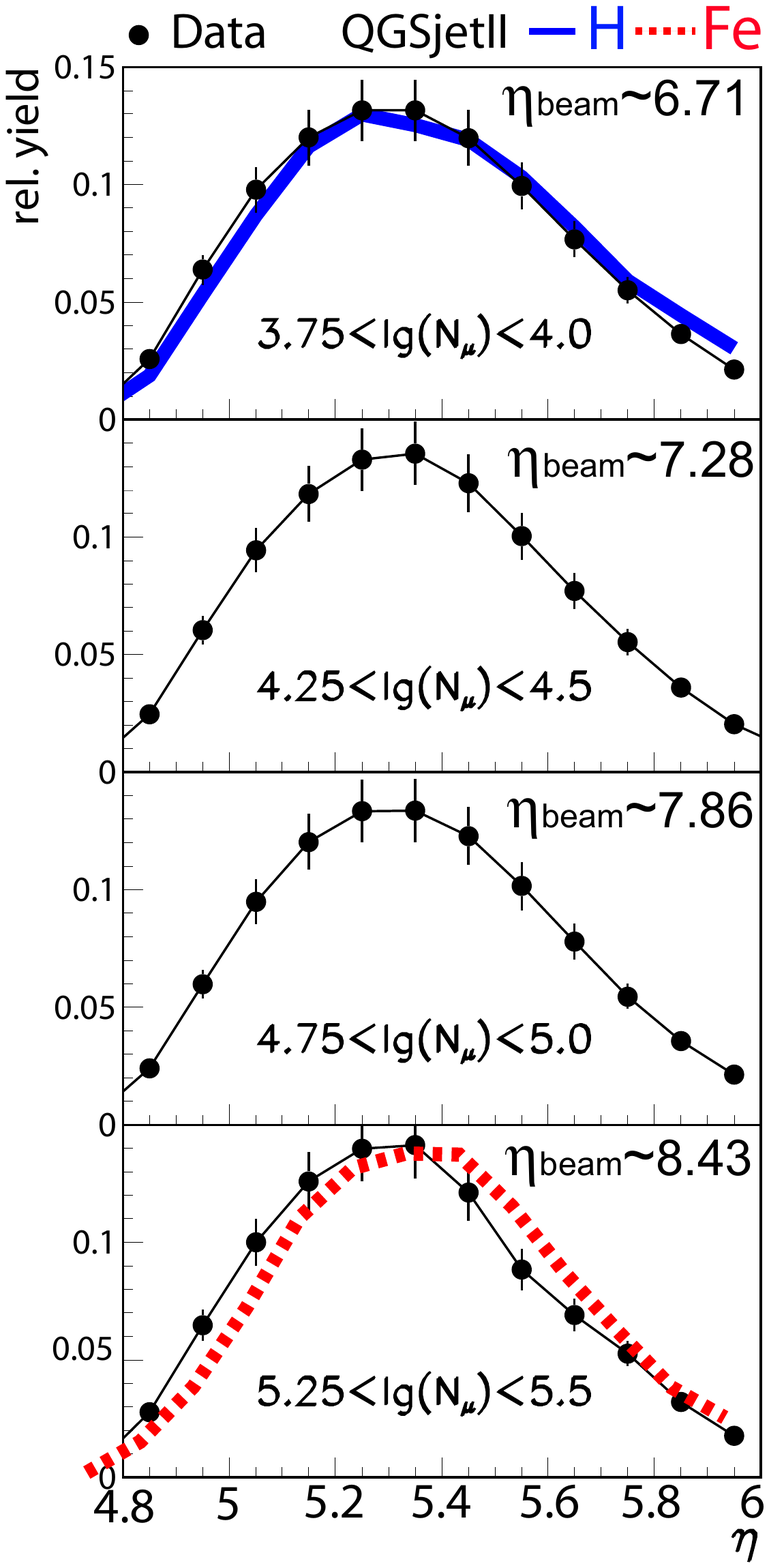}
\end{center}
\vspace{-1.5cm}  \caption{ Pseudorapidity spectra (normalized to
integral yield equal to one) for muon production height $h_{\mu}$
larger than 15 km and 80~-~120 m core distance range and for
shower direction $\Theta=0^{o}-18^{o}$. CORSIKA simulations based
on QGSjetII + FLUKA2002.4 are given for Hydrogen (H) and Iron (Fe)
CR primaries. Lines connect points.} \label{fig3}
\end{figure}


Fig. 2 shows strong reduction of muons from above 15 km for
'light' primary CR particles with respect to Monte-Carlo
especially for the highest energy interval, which corresponds to a
CM energy of $\sim 8~ TeV$. Standard Monte-Carlo (see above)
predicts for a nucleon-nucleon collision charged particle
multiplicities $<N_{ch}>$ $\sim 40~(2TeV)- 70~(8TeV)$. As a
consequence, a fraction out of $\sim 70 \times 200 GeV$ parent
pions are missing. Very recent results from LHC at $7~TeV$ report
also ~\cite{engel} smaller $<N_{ch}>$ at midrapitity compared to
QGSjetII. It would be worthwhile to study to which extent the
simulated distributions in Fig. 2 above $\sim 15~km$ scale with
the inelastic cross sections of H and Fe primary masses. At
$10^{16}$ eV on Nitrogen $\sigma_{inel}^{H}$ amounts to $\sim
400~mb$ and $\sigma_{inel}^{Fe}$ is about 5 times larger. The
ratio of charged particle multiplicities
$<N_{ch}^{Fe}>/<N_{ch}^{H}>$ amounts to about 12.  Both factors
lead to the difference between H and Fe seen in muons in Fig. 2 at
high altitude. Therefore, we assume that about one tenth out of 70
pions lead to an invisible energy fraction corresponding to $7
\times 200 GeV$ ($\sim~1 TeV$) of the incident energy of $\sim
10^{4.5}$~TeV. For lower production height a regular shower
development is taking over and described by the low energy
interaction model FLUKA2002.4. The slower development of showers
seen in the 'light' data compared to simulated H, points in a
classical picture to more $\pi$ decays in the simulations or more
$\mu$ decays in the data, however, can not be reached by
decreasing $\sigma_{inel}$ in MC in the range as discussed in
~\cite{rulrich}.

The deviation from standard high energy Monte-Carlo is strongest
for the 'light' CR primaries for which the quoted $\sqrt S_{NN}$
is around 8~TeV.  The observation that 'heavy' primaries show less
deviation from the Fe prediction, points to a possible threshold
effect for the 'light' CR primaries.

\section{Muon Pseudorapidity}

 To investigate the deviation from the QGSjetII
simulations in Fig. 2~above 15 km especially for the highest
energy bin, pseudorapidity spectra for muon production height
$h_{\mu}$ larger than 15 km have been calculated based on the
$\rho$ and $\tau$ angles as introduced in ~\cite{zabier}.
\begin{equation}
 \eta\simeq ln(2p_{||}/p_{T}) \simeq -ln(\sqrt{\rho^{2} + \tau^{2}}/2)
\end{equation}

The pseudorapidity distributions are taken in the same $N_{\mu}$
size bins. The little variation of the $\eta$ distributions is to
some extent due to our analysis window (15~-~20~km). With our
analysis window we filter out pions around their first
interactions. Only very few 'prompt' muons will be present from
the very first interaction. We have to keep in mind that after a
definite number of interactions pion's probability of decay
exceeds that of arriving at the next interaction level. One should
note, that there is only little difference between $\eta$(pion)
and $\eta$(muon) ~\cite{zabier}. From independent studies with
accelerators the effective beam energy for charged particle
production (inelasticity $K \simeq 0.3$) is quoted to be about
$0.3 \times E_{0}$ ~\cite{basile}. Effective beam rapidities
$\eta_{beam}$ are quoted in the Fig. 3 assuming A=1 CR primary
mass interacting in the atmosphere.

Good agreement for Hydrogen below the 'knee' and for Iron above
the 'knee' is observed. Comparison of $\eta$ distribution for
'light' CR primaries and simulated H in the highest energy bin
suffers from limited statistics.

For the understanding of the $\eta_{peak}$ positions far away from
$\eta_{beam}$ the following considerations may apply. When taking
into account that the CR mass increases up to mass $A\sim56$, beam
rapidities are modified by $-0.5 lnA$, leading to very similar
$\tilde\eta_{beam}$ values. While the total energy increases from
$10^{15}-10^{16.5}$ eV we know ~\cite{doll}~\cite{ulrich}~(in the
frame of QGSjet) that the mean mass of the CR primaries increases
from $<A>\sim 4~to~<A>\sim 56$. Therefore, correcting the quoted
$\eta_{beam}$ gives the $\tilde\eta_{beam}$ values.

 However these $\tilde\eta_{beam}$ values exhibit a
further shift to the $\eta_{peak}$ values possibly due to
geometric scaling ~\cite{wolschin}.
\begin{equation}
 \eta_{peak}=(1/(1+\lambda)) \times (\tilde\eta_{beam}-lnA^{1/6}),~~~\lambda\sim 0.2
\end{equation}

The $\eta$ distributions are described by Monte-Carlo. Therefore,
Monte-Carlo are able to cover the $p_{\|}$ and $p_{t}$ for the
muons which stem mostly from charged pions. Pions are considered
to dominate at midrapidity, but here we deal with the pions in the
fragmentation region which deliver muons conserving the rapidity
of the parent mesons. With respect to the effect of missing muons
at the highest energy, representing a separate feature, the
comparison with other type of high energy interaction model would
be of interest. If this effect is real, it should occur again at
$\sim 10^{18}$ eV for 'heavy' CR particles.

\section{Conclusions}

Muon tracking allows to investigate $h_{\mu}$ and $\eta$ . Future
analysis of other shower angle bins and a larger and improved
quality data sample will provide a more detailed information on
the nature of high energy shower muons. Also muon multiplicities
provide valuable parameters to support the study of the relative
contributions of different primary cosmic ray particles. A natural
extension towards even larger shower energies is provided by
KASCADE-Grande ~\cite{navarra}.



\bigskip 
\begin{acknowledgments}
The KASCADE-Grande experiment is supported by the BMBF of Germany,
the MIUR and INAF of Italy, the Polish Ministry of Science and
Higher Education (Grant for 2009-2011),PPP-DAAD Project for
2009-2010, and the Romanian Authority for Scientific Research
CNCSIS-UEFISCSU (Grant PNII-IDEI no.461/2009 and project PN09 37
01 05.
\end{acknowledgments}

\bigskip 

\end{document}